\documentclass[11pt]{article}
\usepackage[reqno]{amsmath} 
\usepackage{amssymb,theorem,enumerate,verbatim}

\theoremstyle{change}
{\theorembodyfont{\slshape}
\newtheorem{theorem}{Theorem.}[section]
\newtheorem{lemma}[theorem]{Lemma.}
\newtheorem{corollary}[theorem]{Corollary.}

}

\def\sqr#1#2{{\vbox{\hrule height.#2pt
    \hbox{\vrule width.#2pt height#1pt \kern#1pt
        \vrule width.#2pt}\hrule height.#2pt}}}
\def\eqed{\sqr53}
\def\qed{%
    \ifmmode\eqno\eqed
    \else\nobreak\ \hfill\eqed\medbreak\fi}
    
\newenvironment{proof}{\noindent{\sl Proof. }}{\qed}

\newcommand\abs[1]{\left|#1\right|}
\newcommand\charac[1]{\chi\left(#1\right)}

\newcommand\itr[1]{i^{\tr{#1}}}
\newcommand\lb{\langle}
\newcommand\rb{\rangle}
\newcommand\N{N}
\newcommand\omtr[1]{\om^{\tr{#1}}}
\newcommand\scr[1]{{\mathcal #1}}
\newcommand\snv{^{(-)}}
\newcommand\snt{^{(-)T}}

\newcommand\zz[1]{\ints_{#1}}
\newcommand\ip[2]{\left\langle#1,#2\right\rangle}
\newcommand\al{\alpha}

\newcommand\Ga{\Gamma}
\newcommand\la{\lambda}
\newcommand\om{\omega}
\newcommand\Om{\Omega}

\newcommand\Sg{\Sigma}
\renewcommand\th{\theta} 

\newcommand\cx{{\mathbb C}}
\newcommand\re{{\mathbb R}}
\newcommand\fld{{\mathbb F}}
\newcommand\ints{{\mathbb Z}}
\newcommand\diff{\mathbin{\mkern-1.5mu\setminus\mkern-1.5mu}}
\newcommand\inv{^{-1}}
\newcommand\res{\mathbin{\mkern-2.0mu\restriction\mkern-2.0mu}}
\newcommand\sbs{\subseteq}
\newcommand\seq[3]{#1_{#2},\ldots,#1_{#3}}

\newcommand\opk[1]{\mathop{\rm{#1}}\nolimits}

\newcommand\rk{\opk{rk}}
\newcommand\tr{\opk{tr}}

\newcommand\spn{\opk{span}}

\title{Equiangular lines, mutually unbiased bases, and spin models}

\author{Chris Godsil\footnote{email: cgodsil@uwaterloo.ca}\; and Aidan Roy\footnote{email: aproy@uwaterloo.ca. Research supported by NSERC for both authors.}\\
Department of Combinatorics and Optimization \\
University of Waterloo \\
Waterloo, ON N2L 3G1, Canada}

\begin{document}
\maketitle

\begin{abstract}
We use difference sets to construct interesting sets of lines in complex space.
Using $(v,k,1)$-difference sets, we obtain $k^2-k+1$ equiangular lines in $\cx^k$ when $k-1$ is a prime power.  Using semiregular relative difference sets with parameters $(k,n,k,\la)$ we construct sets of $n+1$ mutually unbiased bases in $\cx^k$. We show how to construct these difference sets from commutative semifields and that several known maximal sets of mutually unbiased bases can be obtained in this way, resolving a conjecture about the monomiality of maximal sets. We also relate mutually unbiased bases to spin models.
\end{abstract}

\section{Introduction}

Work in quantum computing has led to renewed interest in certain special sets of lines in
complex space.  A set of $m$ lines in $\cx^d$ spanned by unit vectors $\seq z1m$ is \textsl{equiangular} if there is a constant $a$ such that
\[
|\ip{z_i}{z_j}| = a.
\]
A pair of bases $\seq x1d$ and $\seq y1d$ in $\cx^d$ is \textsl{mutually unbiased} if they 
are both orthonormal and there is a constant $a$ such that
\[
|\ip{x_i}{y_j}| = a
\]
for all $i$ and $j$ when $i\ne j$.  We can view a set of $n$ pairwise mutually unbiased bases in $\cx^d$ as
a set of lines in $n$ groups of size $d$ such that distinct lines in the same group are orthogonal, and if $x$ and $y$ are unit vectors spanning lines in different groups, then
\[
|\ip xy| =a.
\]

It is known that a set of equiangular lines in $\cx^d$ has size at most $d^2$, and that a set mutually unbiased bases contains at most $d+1$ bases.  In the latter case it is known that equality holds if $d$ is a prime power.  Sets of equiangular lines of size $d^2$ are known when $d\in\{2,3,4,5,6,7,8,19\}$ and there is numerical evidence to support the conjecture that such sets exist for all $d$ \cite{ren1}.

In this paper we offer some combinatorial constructions of these types of lines.  Starting with a certain type of difference set in an abelian group, we construct lines from the characters of the group restricted to the difference set. In the case of equiangular lines, a $(v,k,1)$-difference set gives rises to a set of $k^2-k+1$ lines in $\cx^k$ due to K\"onig \cite{kon1}. In the case of mutually unbiased bases, starting with a suitable finite commmutative semifield we find relative difference sets for a group of automorphisms, which give rise to a class of mutually unbiased bases first discovered by Calderbank, Cameron, Kantor and Seidel \cite{ccks}. The procedure results in maximal sets of bases in $\cx^d$ for any prime power $d$.  

We also develop some theory related to these objects.  We present a version of Hoggar's set of 64 equiangular lines in $\cx^8$ \cite{hog1}.  The known constructions for $d^2$ equiangular lines in $\cx^d$ involve taking the action of a group of $d^2$ matrices on a single line in $\cx^d$.  This is also the case for Hoggar, but the group of matrices is not the usual one. We show that the most obvious generalization of this construction does not work in higher dimensions.

Next, we consider equivalence of mutually unbiased bases. Most (but not all) of the known maximal sets are equivalent. Our construction produces several inequivalent sets, which are equivalent to some of those of Calderbank, Cameron, Kantor and Seidel \cite{ccks}; in fact all known maximal sets are encapsulated in their work. We resolve a conjecture of Boykin, Sitharam, Tiep, and Wocjan \cite{bstw}: all known constructions are monomial.

Finally, our construction of sets of lines can be expressed naturally using type-II matrices.  We show that any spin model yields a set of three mutually unbiased bases.

The authors thank Martin R\"otteler for many valuable discussions on the content of this paper.

\section{Difference Sets}\label{sec:eals}

Let $G$ be a group of size $v$.  We work in the complex group algebra $\cx[G]$, which enables us to
identify a subset $S$ of $G$ with the formal sum
\[
\sum_{g\in S}g.
\]
If $\psi$ is a complex-valued function on $G$ and $S\sbs G$, then
\[
\psi(S) :=\sum_{g\in S}\psi(g).
\]
Also
\[
S\inv := \sum_{g\in S}g\inv.
\]

Denote the identity of $G$ in $\cx[G]$ by $1_G$. A subset $D$ of $G$ is a \textsl{$(v,k,\la)$-difference set} if
\[
DD\inv =k1_G+\la(G\diff \{1_G\}).
\]

\begin{theorem}
\label{thm:eal}
The existence of a $(v,k,1)$-difference set in an abelian group implies the existence of a set of $v$ equiangular lines in $\cx^k$.
\end{theorem}

\begin{proof}
Suppose $D$ is such a $(v,k,1)$-difference set in an abelian group $G$, so $v = k^2-k+1$. We construct equiangular lines from the characters of $G$ restricted to $D$. Consider the inner product of two characters:
\begin{align*}
\ip{\chi_a\res D}{\chi_b\res D} & = \sum_{d \in D} \chi_a(d)\overline{\chi_b(d)} \\
	& = \sum_{d \in D} \chi_{ab^{-1}}(d) \\
	& = \chi_{ab^{-1}}(D).
\end{align*}
In particular, $\chi_{ab^{-1}} = \chi_1$, the trivial character, when $a = b$. But the absolute value of $\chi(D)$ satisfies
\begin{align*}
|\chi(D)|^2 & = \chi(D)\overline{\chi(D)} \\
& = \chi(D)\chi(D^{-1}) \\
& = k + \chi(G\diff\{1\}),
\end{align*}
where $\chi(G\diff\{1\})$ is either $v-1$ or $-1$, depending on whether or not $\chi = \chi_1$. So,
\[
\abs{\ip{\chi_a\res D}{\chi_b\res D}}^2 
= \begin{cases}
	k^2, & \mathrm{if\ }\chi_a = \chi_b; \\
	k-1, & \text{ otherwise. }
\end{cases}
\]
Hence after normalizing, the characters are equiangular.
\end{proof}

A $(v,k,1)$-difference set exists if and only if there is a projective plane on $v$ points, with an abelian group of collineations acting regularly on its point set.  Such difference sets have received considerable attention and are known to exist when $k=q+1$, where $q$ is a prime power.  (For more details see \cite[Theorem VI.1.9]{BJL}.) These sets of equiangular lines were first discovered by K\"onig \cite{kon1} and then rediscovered by Xia, Zhou, and Giannakis \cite{xzg}. 

The given $k^2 -k + 1$ lines derived from the characters are {\sl flat}: all entries have the same absolute value. This set is in fact maximal with this property.

\begin{lemma}
There are at most $k^2 - k + 1$ flat equiangular lines in $\cx^k$.
\end{lemma}

\begin{proof}
Let $x_1,\ldots,x_m$ be a set of flat equiangular lines in $\cx^k$, where $|\lb x_i,x_j \rb|^2 = \al$ for $i \neq j$ and each co-ordinate of $x_i$ has absolute value $1/\sqrt{k}$. Consider the Gram matrix of
\[
\Om := \{x_1x_1^*,\ldots,x_mx_m^*\} \cup \{e_1e_1^*,\ldots,e_ke_k^*\};
\]
that is, the matrix whose rows and column are indexed by $\Om$, such that for $uu^*$ and $vv^*$ in $\Om$,
\[
G_{uu^*,vv^*} = \lb uu^*,vv^* \rb = \tr(uu^*vv^*) = |\lb u,v \rb|^2.
\]
This matrix has the form
\[
G = \left( \begin{matrix}
\al J + (1-\al)I & \frac{1}{k}J \\
\frac{1}{k}J & I \\
\end{matrix} \right),
\]
where the first block of $m$ rows and columns is indexed by $x_i$ and the last block of size $k$ is indexed by $e_j$. Also $J$ refers to an all-ones block, not necessarily square. Using elementary row operations, $G$ is row-equivalent to 
\[
G' = \left( \begin{matrix}
(\al - \frac{1}{k})J + (1-\al)I & 0 \\
\frac{1}{k} J & I \\
\end{matrix} \right).
\]
From the eigenvalues of $I$ and $J$ we can determine the rank of $G$. The {\sl relative bound} for $m$ lines in $\cx^k$ (see \cite{dgs}) states that 
\begin{equation}
m \leq \frac{k - k\al}{1 - k\al}.
\label{relbound}
\end{equation}
In fact, $G$ has rank $m+k-1$ if and only if \eqref{relbound} is satisfied with equality; otherwise, it has full rank $m+k$. In either case, the rank is at least $m+k-1$. Moreover, the rank of the Gram matrix is the dimension of the span of $\Om$. The matrices of $\Om$ are Hermitian, so their span (over $\re$) is at most $k^2$. Thus
\[
m+k-1 \leq \rk(G) = \dim(\spn(\Om)) \leq k^2,
\]
and the result follows.
\end{proof}

\section{Hoggar's Construction}
\label{sec:hog}

In \cite{hog1} Hoggar constructed a set of $64$ equiangular lines in $\cx^8$. We describe these lines using a group of $64$ unitary matrices acting on a single vector. Then, we show that $d^2$ equiangular lines in $\cx^d$ can only be constructed with this particular class of matrices when $d \in \{2,8\}$.

Let $X$, $Y$, and $Z$ be Pauli matrices, namely
\[
X = \left(\begin{matrix}
0 & 1 \\
1 & 0 \\
\end{matrix}\right), \; 
Z = \left(\begin{matrix}
1 & 0 \\
0 & -1 \\
\end{matrix}\right), \;
Y = XZ.
\]
Then $\lb X,Z \rb / \lb -I \rb$ is a group of order $4$, sometimes known as the \textsl{ Pauli group}. We are interested in the absolute value of angles between lines under the action of this group, so we can ignore the modulus of $-I$ and represent the Pauli group by the matrices $\{I,X,Y,Z\}$. We construct equiangular lines by applying a tensor product of these matrices to a fixed vector. Let
\[
G = \{I,X,Y,Z\}^{\otimes 3}. 
\]
Then modulo $-I$, we have a group with $64$ elements, each of which is an $8 \times 8$ matrix over $\cx$. Let 
\[
r = \sqrt{2}, \; s = \frac{1+i}{\sqrt{2}}, \; t = \frac{1-i}{\sqrt{2}},
\]
and let 
\[
v = (0,0,s,t,s,-s,0,r).
\]
Then
\[
\{ Av \mid A \in G \}
\]
is a set of equiangular lines, equivalent to that of Hoggar. 

Now consider $v \in C^d$ under the action of the Pauli group for any $d = 2^k$. Let 
\[
G_k = \{I,X,Y,Z\}^{\otimes k}
\]
(again mod $-I$), and let $v = (v_1,v_2,\ldots, v_d)$. 

\begin{lemma}
The lines
\[
\{ Av \mid A \in G_k \}
\]
can only be equiangular for $k = 1$ or $k = 3$.
\end{lemma}

\begin{proof}
We establish a system of equations with the coordinates of $v$ as variables, and show solutions can only exist for the given $k$. Let $\al_i$ denote $v_i^*v_i$. Since $v^*v = 1$, we have
\[
\al_1+\ldots+\al_d = 1.
\]
Similarly, from $|v^*(I \otimes \ldots \otimes I \otimes Z)v| = \frac{1}{\sqrt{d+1}}$, we have
\[
\al_1-\al_2+\ldots+\al_{d-1}-\al_d = \pm \frac{1}{\sqrt{d+1}}.
\]
(Since each $\al_i$ is real, the right-hand side must also be real.) More generally, let $\al = (\al_1,\ldots,\al_d)$. Then by considering $v^*Av$ for $A \in \{I,Z\}^{\otimes k}$, we get the system of equations
\[
H \al = \frac{1}{\sqrt{d+1}}\left(\begin{matrix}
\sqrt{d+1} \\
\pm 1 \\
\vdots \\
\pm 1
\end{matrix} \right),
\]
where $H$ is a $d \times d$ Hadamard matrix:
\[
H = \left(\begin{matrix} 
1 & 1 \\ 
1 & -1 
\end{matrix} \right)^{\otimes k}.
\]
Since $H^{-1} = \frac{1}{d}H$, this system is easily solved for $\al$:
\begin{equation}
\al_i = \frac{\sqrt{d+1} + l_i}{d \sqrt{d+1}},
\label{eqn:ali}
\end{equation}
for some odd integer $l_i$. Next, consider terms of the form $f_i = v_i^*v_{i+1}$. Again the angles between lines lead to a system of equations. Let $X_k = I \otimes \ldots \otimes I \otimes X$. Since $|v^*X_kv| = \frac{1}{\sqrt{d+1}}$, we have
\[
f_1 + f_1^* + \ldots + f_{d-1} + f_{d-1}^* = \pm \frac{1}{\sqrt{d+1}}.
\]
(Again, since $f_i + f_i^*$ is real, the right-hand side is also real.) From $|v^*X_k(I \otimes \ldots \otimes I \otimes Z)v| = \frac{1}{\sqrt{d+1}}$, we have
\[
f_1 - f_1^* + \ldots + f_{d/2} - f_{d/2}^* = \pm \frac{\om}{\sqrt{d+1}},
\]
where $\om = \sqrt{-1}$. (Since $f_i - f_i^*$ is purely imaginary, the right side is purely imaginary.) More generally, letting $f = (f_1, f_1^*, \ldots, f_{d-1}, f_{d-1}^*)$, and considering $v^*X_kAv$ for $A \in \{I,Z\}^{\otimes k}$, it follows that 
\[
H f = \frac{1}{\sqrt{d+1}}\left(\begin{matrix}
\pm 1 \\
\pm \om \\
\vdots \\
\pm 1 \\
\pm \om
\end{matrix} \right).
\]
The solutions in $f$ are of the form
\[
f_i \in \frac{\pm \{0,2,4,\ldots\} \pm \{0,2,4,\ldots\}\om}{d \sqrt{d+1}}.
\]
Thus,
\[
f_if_i^* = \frac{m}{d^2(d+1)},
\]
for some integer $m$; that is, $f_if_i^*$ is rational. However, there is nothing special about $f_i$, the ``cross term'' of $v_i$ and $v_{i+1}$. For any $i \neq j$, let $g = v_i^*v_j$, and let $M \in \{I,X\}^{\otimes k}$ be the permutation matrix from the Pauli group that takes coordinate $i$ to $j$ (and $j$ to $i$) on $v$. Then by considering $v^*MAv$, $A \in \{I,Z\}^{\otimes k}$, we similarly get that $gg^*$ is rational.

Lastly, note that if $g = v_i^*v_j$, then $gg^* = \al_i\al_j$, and this product is rational. However, from formula \eqref{eqn:ali} for $\al_i$,
\[
\al_i\al_j = \frac{d+1 + l_il_j + (l_i+l_j)\sqrt{d+1}}{d^2(d+1)},
\]
which is rational if and only if $\sqrt{d+1}$ is rational or $l_i = -l_j$. If $l_i = -l_j$ for all $i \neq j$, then there are only two possible indices of $i$ and $j$, so $d = 2$. On the other hand, $\sqrt{2^k+1}$ is rational only if $k = 3$. We conclude that the lines can only be equiangular for $d \in \{2,8\}$. 
\end{proof}

\section{Mutually Unbiased Bases}
\label{sec:mubs}

Let $G$ be a group and $N$ a normal subgroup of $G$.  A subset $D$ of $G$ is a 
\textsl{relative difference set} if there is an integer $\la$ such that
\[
DD\inv =|D|1_G+\la(G\diff N).
\]
It is customary to assume that  $n=|N|$, $|G| = mn$, and $|D| = k$.  With these 
conventions we  say $D$ is a \textsl{$(m,n,k,\la)$ relative difference set}; it is 
\textsl{semi-regular} if $m = k$.

\begin{theorem}
\label{thm:mub}
The existence of a semi-regular $(k,n,k,\la)$-relative difference set in
an abelian group implies the existence of a set of $n+1$ mutually unbiased
bases of $\cx^k$. 
\end{theorem}

\begin{proof}
As with difference sets, we construct mutually orthogonal bases from the characters of $G$ restricted to the set $D$.

The characters $G^*$ of $G$ form a group, as do the characters of $G/\N$. Moreover, every character of $G/\N$ induces a character of $G$ which is on constant on the cosets of $\N$, and these characters form a subgroup of size $k$. Denote this subgroup by $H^*$.

Define basis of $B_i$ of $\cx^k$ to be the $i$-th coset of $H^*$ (restricted to $D$). As with difference sets, the inner product of two characters is:
\begin{align*}
\ip{\chi_a\res D}{\chi_b\res D} & = \sum_{d \in D} \chi_a(d)\overline{\chi_b(d)} \\
& = \chi_{ab^{-1}}(D).
\end{align*}
It is easy to show (see \cite[Lemma 10.9]{BJL} for example) that $\chi(D)$ has absolute value $k^2$, $0$, or $k$, depending on if $\chi$ is in $\{\chi_1\}$, $H^*\diff\{\chi_1\}$ or $G^*\diff H^*$ respectively.  Normalizing each character,
\[
\abs{\ip{\chi_a\res D}{\chi_b\res D}} 
= \begin{cases}
	\sqrt{k}, & \chi_a = \chi_b; \\
	0, & \chi_a \neq \chi_b \text{ but } \chi_a\chi_b^{-1} \in H^*; \\
	1, & \text{ otherwise. }
\end{cases}
\]
Hence the $n$ normalized bases $B_i$ are orthogonal and mutually unbiased. 
Since every entry of every basis element has norm $1/\sqrt{k}$, the standard basis is also unbiased with each $B_i$.
\end{proof}

With some additional assumptions on the
bases involved, this construction can be reversed.
Suppose $u$ is an vector in a basis $B$ which is mutually
unbiased with the standard basis. Then each entry of $u$ has norm
$1/\sqrt{k}$. By multiplying $u$ by $\sqrt{k}$, each entry has norm $1$.
We will call this operation {\sl Schur-normalization}. If we
Schur-normalize all the vectors from an entire set of mutually unbiased
bases, then vectors $u_1$ and $u_2$ from bases $B_1$ and $B_2$ satisfy
\[
|\ip{u_1}{u_2}| = \begin{cases}
k, & u_1 = u_2; \\
0, & u_1 \neq u_2 \text{ but } B_1 = B_2; \\
\sqrt{k}, & \text{ otherwise. }
\end{cases}
\]
Note that a necessary condition for a collection of vectors to form a group
under Schur multiplication is that each vector in the collection must be Schur-normalized.

\begin{corollary}
Let $B_1, \ldots, B_n$ be mutually unbiased bases of $\cx^k$, each
mutually unbiased with the standard basis.  If the vectors of
$B_1 \cup \ldots \cup B_n$ form a group (of size $nk$) with respect to
Schur multiplication, then there exists a semiregular relative difference
set with parameters $(k,n,k,\la)$. 
\end{corollary}

Semiregular relative difference sets are closely related to antipodal covering graphs: every $(k,n,k,\la)$-relative difference set is equivalent to an antipodal distance-regular $n$-fold cover of $K_{k,k}$ with an automorphism group acting regularly on each colour class. See Godsil \cite{god8} for more details on covers of complete bipartite graphs.

\section{Semifields}
\label{sec:sfields}

Semiregular relative difference sets are not easy to find.  We intend to construct them from commutative semifields, and so it behooves us to describe these first.

Roughly speaking, a semifield is a field where multiplication need not be associative.
More formally, a \textsl{finite semifield} is a finite set $E$ with two operations, addition $+$ and multiplication $\circ$, such that
\begin{enumerate}[(a)]
\item
$E$ is an abelian group under addition, with identity $0$.
\item
If $x\circ y=0$, then $x=0$ or $y=0$.
\item
$x\circ(y+z)=x\circ y +x\circ z$ and $(y+z)\circ x=y\circ x+z\circ x$.
\item
There is a multiplicative identity $1$.
\end{enumerate}
Every finite field is a semifield.  The \textsl{right nucleus} of $E$ is the set
\[
\{x: a\circ (b\circ x)=(a\circ b)\circ x\}.
\]
This contains the additive subgroup of $E$ generated by $1$, which is a field of prime
order.  It can be shown that $E$ is a vector space over this field, and consequently
a finite semifield has prime power order.  It can also be shown that, if $p$ is prime,
a semifield of order $p$ or $p^2$ is a field.

Using a semifield $E$, we construct an incidence structure as follows.  The point set of
the incidence structure is just $E\times E$; we denote the points by ordered pairs $(x,y)$.
The line set is a second copy of $E\times E$, where we denote a line
by $[a,b]$.  The element $a$ of the line $[a,b]$ is called its \textsl{slope}.  The point 
$(x,y)$ lies on the line $[a,b]$ if and only if
\[
y=a\circ x+b.
\]
If for each $c$ in $E$ we adjoin the line consisting of the points
\[
(c,y),\quad y\in E,
\]
then the resulting incidence structure is the affine plane $AG(2,E)$.

We construct some groups of automorphisms.  If $a,b\in E$ define the map $T_{a,b}$ by
\[
T_{a,b}(x,y):=(x+a,y+b).
\]
It is easy to check that if $(x,y)$ lies on $[u,v]$, then $T_{a,b}(x,y)$ lies on
$[u,v+b-u\circ a]$.  Therefore $T_{a,b}$ is an automorphism and the set
\[
T := \{T_{a,b}:a,b\in  E\}
\]
is an abelian group that acts transitively on points, and with each parallel 
class of lines forming an orbit of lines.  (A parallel class is the set of lines with a
given slope.)

We similarly define a map $S_{u,v}$ on lines by
\[
S_{u,v}([r,s]) := [r+u,s+v].
\]
It is not hard to show that
\[
S_{u,v}(x,y) =(x,y+u\circ x+v).
\]
Therefore $S_{u,v}$ is an automorphism and
\[
S :=\{S_{u,v}: u,v\in E\}
\]
is an abelian group that acts transitively on the lines and has the point sets of
the lines of infinite slope as its point orbits.

Now define $H_{u,b}$ by
\[
H_{u,b} := T_{u,b}S_{u,0}.
\]
Then
\[
H_{u,b}H_{v,d} =H_{u+v,b+d+u\circ v}.
\]
Given this it is not hard to show that 
\[
H :=\{H_{u,b}:u,v\in E\}
\]
is a group and that if $E$ is commutative, then $H$ is commutative.
We also find that $H$ acts transitively on points and lines. This result is originally due to Hughes \cite{hug1} in 1956.

\begin{theorem}
Let $E$ be a finite commutative semifield of order $q$.  Then the group $H$ is abelian 
with order $q^2$, and the subset
\[
H_0 = \{H_{u,0}: u\in E\}
\]
is a relative difference set in $H$ with parameters $(q,q,q,1)$. \qed
\end{theorem}

\begin{corollary}
Let $E$ be a finite commutative semifield of order $q$.  Then the the characters of $H$ restricted to $H_0$, together with the standard basis, are a set of $q+1$ mutually unbiased bases in $\cx^q$. \qed
\label{cor:semi}
\end{corollary}

For a survey of finite semifields, see Cordero and Wene \cite{cw1}. We now construct mutually unbiased bases explicitly using the characters of $H$. 

\begin{lemma}
When $q = p^n$ is odd, let $\om$ be a primitive $p$-th root of unity, and let $\ip{a}{x}$ denote the scalar product from $E \times E$ to $GF(p)$. Then
\[
\phi_{ab}(H_{x,y}) = \om^{\ip{2a}{x}+ \ip{b}{2y-x \circ x}}
\]
is a character of $H$.
\label{lem:oddsemi}
\end{lemma}

\begin{proof}
\begin{align*}
\phi_{ab}(H_{x,y})\phi_{ab}(H_{w,z}) & =  \om^{\ip{2a}{x} + \ip{b}{2y-x \circ x}} \om^{\ip{2a}{w} + \ip{b}{2z-w \circ w}} \\
& = \om^{\ip{2a}{x+w} + \ip{b}{2(y+z+x \circ w) - (x+w)^{\circ 2}}}  \\
& = \phi_{ab}(H_{x+w,y+z+x \circ w}).
\end{align*}
\end{proof}

When $q = 2^n$ is even, we need more structure. Let $\{e_1, \ldots, e_n\}$ be a basis for $E$ over $GF(2)$, let $\{\widehat{e}_1, \ldots, \widehat{e}_n\}$ be a basis for $R$, a free module over $\zz{4}$. For each $x = \sum x_ie_i$ in $E$, $x_i \in \zz{2}$, embed $x$ in $R$ as
\[
x \mapsto \widehat{x} = \sum_{i=1}^n x_i\widehat{e}_i.
\]
Since $x_i \in \{0,1\}$, any element of $R$ can be written uniquely in the form $\widehat{x} + 2\widehat{y}$, with $x,y \in E$. This map is not an additive homomorphism, but it does preserve addition mod $2$: for any $x$ and $y$ in $E$,
\[
2(\widehat{x} + \widehat{y}) = 2\widehat{x+y}.
\]
Define multiplication on $R$ as follows: let 
\[
\widehat{e}_i\widehat{e}_j = \widehat{e_i \circ e_j}
\]
for basis elements $\widehat{e}_i$ and $\widehat{e}_j$, and extend linearly to all of $R$. Then multiplication distributes over addition, and the embedding preserves multiplication mod $2$: 
\[
2\widehat{x}\widehat{y} = 2\widehat{x \circ y}.
\]
Finally, note that since $\widehat{x} + \widehat{y} = \widehat{x+y} + 2\widehat{z}$ for some $z \in E$, we have
\[
(\widehat{x} + \widehat{y})^2 = (\widehat{x+y} + 2\widehat{z})^2 = (\widehat{x+y})^2.
\]
With these properties, the proof of the following is the same as Lemma \ref{lem:oddsemi}.
\begin{lemma}
Let $i$ denote a primitive $4$-th root of unity, and let $\ip{u}{v}$ denote the scalar product from $R \times R$ to $\zz{4}$. Then for $a,b$ in $E$,
\[
\phi_{ab}(H_{x,y}) = \om^{\ip{2\widehat{a}}{\widehat{x}}+ \ip{\widehat{b}}{2\widehat{y}-\widehat{x}^2}}
\]
is a character of $H$. \qed
\label{lem:evensemi}
\end{lemma}

In the special case when $E$ is in fact a field, the characters of $H$ are simpler. When $q = p^n$ is odd, using the trace function $\tr: E \rightarrow GF(p)$, the characters may be written as
\begin{equation}
\phi_{ab}(H_{x,y}) = \om^{\tr(2ax + b(2y-x^2))}.
\label{eqn:oddfield}
\end{equation}

When $q = 2^n$ is even, take $R$ to be the Galois ring $GR(4^n)$ and embed $E$ into the Teichm\"uller set $T$ of $R$. That is, identify each $x \in E$ with the element of $T$ congruent to $x$ mod $2$. Letting $\tr$ denote the Galois ring trace from $R$ to $\zz{4}$, 
\begin{equation}
\phi_{ab}(H_{x,y}) = i^{\tr(2ax + b(2y-x^2))}
\label{eqn:evenfield}
\end{equation}
is a character for all $a,b \in E$. 

These characters of $H$, when restricted to a relative difference set in the group (that is, taking $y = 0$), form set of $q$ mutually unbiased bases in $\cx^q$. As we explain in the next section, the bases from equations \eqref{eqn:oddfield} and \eqref{eqn:evenfield} are equivalent to those of Klappenecker and R\"{o}tteler in \cite{kr1}. 

\section{Equivalence}
\label{sec:equiv}

Equivalence of mutually unbiased bases was introduced by Calderbank, Cameron, Kantor, and 
Seidel \cite{ccks}. Identify a vector in $\cx^n$ with a point in projective space $PG(n-1,\cx)$, so 
that two vectors in $\cx^n$ are considered the same if they span the same $1$-dimensional 
subspace. Two sets of mutually unbiased bases $\{B_0,\ldots,B_d\}$ and 
$\{B_0', \ldots, B_d'\}$ are \textsl{equivalent} if there is a unitary operator $U$ mapping 
the first set of bases to the second set (in no particular order):
\[
\{ U(B_0), \ldots, U(B_d)\} = \{B_0', \ldots, B_d'\}.
\]
Note that $U$ preserves angles between lines: for any two subspaces $\lb x \rb$ and $\lb y \rb$, 
\[
|\lb Ux,Uy \rb| = |\lb x,y \rb|.
\]

In \cite{ccks}, Calderbank et al.~find several inequivalent mutually-unbiased bases 
(which they refer to as \textsl{orthogonal frames}) using symplectic spreads and 
$\zz{4}$-Kerdock codes. In particular, given a symplectic spread $\Sg$, they show how 
to construct a maximal set of mutually unbiased bases $\scr{F}(\Sg)$ (Theorem 5.6 in the
 even case; Theorem 11.4 in the odd case). They then show that two sets of bases 
 $\scr{F}(\Sg_1)$ and $\scr{F}(\Sg_2)$ are equivalent if and only if there is a symplectic transformation sending $\Sg_1$ to $\Sg_2$ (Proposition 5.11 and Corollary 11.6). Using Kantor's result \cite{kan1} on inequivalent symplectic spreads, Calderbank et al. conclude that a large number of inequivalent sets of mutually unbiased bases exist for $\cx^n$ where $n$ is an odd power of $2$. 

Since there is a natural correspondence between semifields and symplectic semifield spreads (see for example Kantor \cite[Proposition 3.8]{kan2}), our mutually unbiased bases in Corollary \ref{cor:semi} are included in those of Calderbank et al.\footnote{See \cite{kan3} for further explanation.} In fact, all known maximal sets fit into that framework. In this section, we show that the constructions of Alltop \cite{all1}, Wootters and Fields \cite{wf1}, Klappenecker and R\"otteler \cite{kr1}, and Bandyopadhyay, Boykin, Roychowdhury, and Vatan \cite{bbrv} are all equivalent and are a special case of the constructions of Calderbank et al. In the process, we show that all known constructions of maximal sets of mutually unbiased bases are ``monomial", answering a conjecture of Boykin, Sitharam, Tiep, and Wocjan \cite{bstw}.

Throughout this section, we express mutually unbiased bases in terms of matrices, where the columns of each $q \times q$ matrix form the bases for $\cx^q$. So, two sets of matrices are equivalent if there is a unitary map taking one set to the other, up to permutations of columns and multiplying any column by an element of $\cx$ of modulus $1$.

Firstly, consider the odd case. The earliest construction was due to Alltop \cite{all1}, although he expressed his result in different terms. Let $\fld$ denote the finite field of order $q$ and characteristic $p$ ($p$ odd). As before, $\tr$ is the $GF(p)$-valued trace on $\fld$ and $\om$ is a primitive $p$-th root of unity.  Define the matrices $A_\al$ by
\[
A_\al := \frac1{\sqrt{q}}\left(\omtr{(x+\al)^3+y(x+\al)}\right)_{x,y}, \quad x,y\in\fld.
\]
If $p > 3$, then $\{A_\al: \al \in \fld\} \cup \{I\}$ is a set of $q+1$ mutually unbiased bases for $\cx^q$. 

The next construction was originally due to Ivanovic \cite{iva1} (in the case of prime dimension) and Wootters and Fields \cite{wf1} (who generalized Ivanovic's work to all prime powers). Define $W_\al$ by
\[
W_\al := \frac1{\sqrt{q}}\left(\omtr{(\al x^2+xy)}\right)_{x,y},\quad x,y\in\fld.
\]
Klappenecker and R\"{o}tteler \cite[Theorem 2]{kr1} gave a simplified proof that the matrices
$W_\al$ together with the identity matrix form a set of mutually unbiased bases.

\begin{lemma}
For $p > 3$,
\[
\{A_\al: \al \in \fld\} \cup \{I\}
\]
is equivalent to 
\[
\{W_\al: \al \in \fld\} \cup \{I\}.
\]
\end{lemma}

\begin{proof}
For convenience, let 
\[
\chi(x) := \omtr{x}.
\]
Multiply each $A_\al$ on the left by the unitary matrix $A_0^*$. Since $A_0^* = A_0^{-1}$, this map takes $A_0$ to $I$ and $I$ to $A_0^*$ (which, after dividing column $x$ by $\omtr{x^3}$, is $W_0$).  In the remaining cases:
\begin{align*}
\left(A_0^*A_\al\right)_{x,y} & = \sum_{z \in \fld} \left(A_0^*\right)_{x,z}\left(A_{\al}\right)_{z,y} \\
& = \frac{1}{q}\sum_{z \in \fld} \charac{-z^3-xz}\charac{(z+\al)^3+y(z+\al)} \\
& = \frac{1}{q}\sum_{z \in \fld} \charac{3\al z^2 + (3\al^2+y-x)z + (\al^3+y\al)}.
\end{align*}
This expression is known as a Weil sum and can be evaluated with the following formula from Lidl and Niederreiter \cite[Theorem 5.33]{lid1}:
\[
\sum_{z \in \fld} \charac{a_2z^2 + a_1z + a_0} = \charac{a_0 - \frac{a_1^2}{4a_2}} \eta(a_2)G(\eta, \chi).
\]
Here $\eta(a_2)$ is the quadratic residue of $a_2$ and $G(\eta, \chi)$ is a Gaussian sum which is independent of $a_0$, $a_1$ and $a_2$. Thus,
\[
\left(A_0^*A_{\al}\right)_{x,y} = \frac{1}{q}\charac{\frac{12\al^4 + 12y\al^2 - (3\al^2+y-x)^2}{12\al}} \eta\left(3\al\right)G.
\]
Now divide each column by its entry in the row $x = 0$, namely $(A_0^*A_{\al})_{0,y}$. (This does not affect the absolute value of the angle between the columns.) Most of the terms cancel. The result is
\begin{align*}
\frac{\left(A_0^*A_{\al}\right)_{x,y}}{\left(A_0^*A_{\al}\right)_{0,y}} & = \charac{\frac{-x^2 +2x(3\al^2+y)}{12\al}} \\
& = \charac{-\frac{1}{12\al}x^2 + \frac{3\al^2+y}{6}x} \\
& = \left(W_{-\frac{1}{12\al}}\right)_{x,\frac{3\al^2+y}{6}}.
\end{align*}
We conclude that pre-multiplying by $A_0^*$ maps $A_\al$ to $W_{-1/12\al}$, up to the column permutation $y \mapsto (3\al^2+y)/6$. Thus the mutually unbiased bases are equivalent. 
\end{proof}

Bandyopadhyay, Boykin, Roychowdhury, and Vatan \cite{bbrv} gave another construction of the same bases. Let $\{e_u\}$ denote the standard basis for $\cx^q$, indexed by the elements of $\fld$. For $a$ in $\fld$, define the following $q \times q$ matrices:
\begin{align*}
X(a): &\;  e_u \mapsto e_{u+a}, \\
Z(a): &\;  e_u \mapsto \om^{\tr(au)}e_u.
\end{align*}

Clearly, the standard basis is a complete set of eigenvectors for $Z(a)$. It is also easy to verify that the vectors
\[
f_u = \sum_{v \in \fld} \om^{\tr(uv)} e_v, \quad u \in \fld
\]
form a complete set of eigenvectors for $X(a)$.

Define a map from $\fld^2$ to the $d \times d$ matrices as follows:
\[
D_{(a,b)} := X(a)Z(b).
\]
Up to a phase, these matrices are sometimes called the \textsl{generalized Pauli matrices}. Each $D_{(a,b)}$ is unitary and monomial, and $\{D_{(a,b)}\}$, modulo scalar multiples of $I$, is isomorphic to $\fld^2$ as a group. Bandyopadhyay et al. partition these matrices into commuting sets and show that the common eigenvectors must form mutually unbiased bases. Those eigenvectors are the bases of Wootters and Fields.

\begin{lemma} Let $a$, $c$, $d$, and $b = 2ac$ be in $\fld$. Then
\begin{equation}
\phi_{c,d} = \sum_{x \in \fld} \om^{\tr(cx^2 + 2dx)} e_x
\label{eqnevec}
\end{equation}
is an eigenvector for $D_{a,b}$. \qed
\label{lem:oddmono}
\end{lemma}

Next, consider the case where $q$ is even. Again, the first construction was due to Wootters and Fields, but Klappenecker and R\"otteler \cite[Theorem 3]{kr1} gave a simpler description. With $q = 2^n$, as before let $T$ denote the Teichm\"uller set of the Galois ring $R = GR(4^n)$ and let $\tr: R \rightarrow \ints_4$ denote the trace.  Define
\[
W_\al :=\left(\itr{(\al+2y)x}\right)_{x,y},\quad x,y\in T.
\]
Note that $\tr(x^2) = \tr(x)$ in $T$, so after some permutation of $\al$ these matrices have the form
\[
W_\al = \left(\itr{\al x^2+2yx}\right)_{x,y},\quad x,y\in T,
\]
which are equivalent to those after Lemma \ref{lem:evensemi}.

Again, Bandyopadhyay et al. constructed the same bases using Pauli matrices. The connection between the bases defined using $R = GR(4^n)$ and the Pauli matrices, defined over $\fld = GF(2^n)$, is the natural mod $2$ mapping. In addition to being a ring homomorphism from $R$ to $\fld$, it is bijection from $T$ to $\fld$.

Recall that $T$ is multiplicatively closed, and any element of $R$ can be written $x + 2y$ for $x,y \in T$. Note that $(x + 2y)^2 = x^2$ is in $T$. Also $(x + y)^2 = x^2 + y^2 + 2xy$, so for any $x$ and $y$ in $T$, $x + y + 2\sqrt{xy}$ is the unique element of $T$ congruent to $x + y$ mod $2$. 

Using the bijection between $T$ and $\fld$, the Pauli matrices are, with $a,u \in T$, 
\begin{align*}
X(a) &: e_u \mapsto e_{u+a+2\sqrt{ua}}, \\
Z(a) &: e_u \mapsto (-1)^{\tr(au)} e_u = i^{\tr(2au)} e_u.
\end{align*}
As in the case of $q$ odd, the eigenvectors of $D_{a,b}$ are the bases described by Klappenecker and R\"otteler.

\begin{lemma} Let $a$, $c$, $d$, and $b = ac$ be in $T$. Then
\[
\phi_{c,d} = \sum_{x \in T} i^{\tr(cx^2 + 2dx)} e_x
\]
is an eigenvector for $D_{a,b}$.
\label{lem:evenmono}
\end{lemma}

\begin{proof}
\begin{align*}
X(a)Z(b)\phi_{c,d} & = \sum_{x \in G} \om^{\tr(cx^2 + 2dx)} X(a)Z(b) e_x \\
& = \sum_{x \in G} i^{\tr(cx^2 + 2dx + 2bx)} e_{x+a+2\sqrt{xa}} \\
& = \sum_{x \in G} i^{\tr(c(x+a)^2 + 2d(x+a) + 2bx - 2cax - ca^2 - 2da)} e_{x+a+2\sqrt{xa}} \\
& = i^{-\tr(ca^2 + 2da)} \sum_{x \in G} i^{\tr(c(x+a)^2 + 2d(x+a))} e_{x+a+2\sqrt{xa}} \\
& = i^{-\tr(ca^2 + 2da)} \phi_{c,d}.
\end{align*}
In the second last line, $(x+a)^2 = (x+a+2\sqrt{xa})^2$ and $2(x+a) = 2(x+a+2\sqrt{xa})$.
\end{proof} 

\section{Monomiality and Nice Error Bases}

In their construction, Bandyopadhyay et al. \cite[Theorems 3.2 \& 3.4]{bbrv} show that any set of mutually unbiased bases $\scr{B} = \{B_1, B_2, \ldots, B_m\}$ in $\cx^n$ is equivalent to a \textsl{maximal commuting basis} of matrices: a collection of $n \times n$ unitary matrices $C = C_1 \cup C_2 \cup \ldots \cup C_m$ such that
\begin{enumerate}[(a)]
\item $|C_i| = n$, 
\item $I \in C_i$, 
\item the matrices of $C_i$ commute, and
\item the matrices of $C$ are pairwise orthogonal with respect to the trace inner product.  
\end{enumerate}
More specifically, basis $B_i$ is the set of common eigenvectors for $C_i$, and we say $\scr{B}$ is obtained by \textsl{partitioning} $C$. For the bases of the previous section, Lemmas \ref{lem:oddmono} and \ref{lem:evenmono} indicate that $C = \{D_{a,b}\}$, the generalized Pauli matrices. 

Boykin, Sitharam, Tiep, and Wocjan \cite{bstw} define a set of mutually unbiased bases to be \textsl{monomial} if it is equivalent to a set of bases in which all of the matrices of $C$ are monomial. Since each $D_{a,b}$ is a monomial matrix, the mutually unbiased bases of Wootters and Fields (or the equivalent reformulations of \cite{kr1} or \cite{bbrv}) are monomial. In fact, this result holds for all of the mutually unbiased bases in Corollary \ref{cor:semi}: the bases are equivalent to those of Calderbank et al., which are monomial by construction.

A set $C$ of $n \times n$ unitary matrices is called a \textsl{nice error basis} if: 
\begin{enumerate}[(a)]
\item the matrices are pairwise orthogonal (with respect to the trace inner product), and 
\item modulo scalar multiples of the identity, $C$ is isomorphic a group of order $n^2$.
\end{enumerate}
Again, the generalized Pauli matrices are the canonical example.  So, the mutually unbiased bases of Wootters and Fields are obtained by partitioning nice error bases.  The bases of Calderbank et al. are also obtained by partitioning the generalized Pauli matrices.  This verifies the conjecture of Boykin, Sitharam, Tiep, and Wocjan \cite[Conjecture 3.4]{bstw}: all maximal constructions of mutually unbiased bases that we know of are both monomial and obtained by partitioning nice error bases.

\section{Spin Models}
\label{sec:spin}

Spin models were introduced by Jones in \cite{jon1}.  Here we show that they can be used to
construct mutually unbiased bases.

The \textsl{Schur product} $M\circ N$ of two $m\times n$ matrices $M$ and $N$ 
is the $m\times n$ matrix such that
\[
(M\circ N)_{i,j} :=M_{i,j}N_{i,j}.
\]
The all-ones matrix $J$ is an identity matrix for the Schur product.
A matrix $M$ has an inverse with respect to the Schur product if and only if all its
entries are non-zero; we denote the Schur inverse of a Schur invertible matrix $M$
by $M\snv$.  Finally a $v\times v$ matrix $M$ is a \textsl{type-II matrix} if it is invertible 
and Schur invertible and
\[
MM\snt =vI.
\]
Recall that matrix is flat if all its entries have the same absolute value.  The following
lemma is easy to prove.

\begin{lemma}
Let $M$ be a square matrix.  Then any two of the following statements imply the third:
\begin{enumerate}[(a)]
\item
$M$ is a type-II matrix.
\item
Some nonzero scalar mutliple of $M$ is unitary.
\item
$M$ is flat. \qed
\end{enumerate}
\end{lemma}

As  a corollary we  note that if the unitary matrix $M$ is unbiased relative to the identity matrix, then it is a flat type-II matrix. These matrices are sometimes called \textsl{complex Hadamard matrices}.

If $M$ is Schur invertible, then $M_{i/j}$ denotes the `Schur ratio'
\[
M_{i/j} := (Me_i)(Me_j)\snv.
\]
A type-II matrix $W$ is a \textsl{spin model} if each of the ratios $W_{i/j}$ is an eigenvector
for $W$.  For example, if
\[
a :=\frac12\left[-v+2\pm\sqrt{v^2-4v}\right]
\]
then $(a-1)I+J$ is a type-II matrix, and 
it is easy to show that it is a spin model (known as the Potts model).
Spin models are interesting because each spin model gives to an invariant of knots
and links.  For example, the Potts model gives rise to the Jones polynomial.  Some spin models also provide sets of mutually unbiased bases, as we now show.

Our next result is a consequence of Lemma 4.3 and Lemma 9.2 from \cite{CGM}.

\begin{lemma}
\label{lem:adadad}
Let $A$ be a type-II matrix of order $n\times n$  and let $D_j$ be the diagonal matrix
with $r$-th diagonal entry equal to the $r$-th entry of the $j$-th column of $\sqrt{n}A\snv$.
If  $A$ is a spin model, then for $j=1,\ldots,n$,
\[
D_jAD_j\inv =A\inv D_j A. \qed
\]
\end{lemma}

If $A$ is unitary and flat, then the diagonal entries of $D_j$ all have norm 1.  From this
it follows that each diagonal entry of $A\inv D_jA$ is equal to $\tr(D_j)$.  
On the other hand, $(D_jAD_j\inv)_{i,i}=A_{i,i}$.  Therefore the diagonal entries
of $A$ are constant, and so each is equal to $1/\sqrt{n}$.  Consequently $\tr(D_j)=1/\sqrt{n}$
for each $j$, which shows that that column sums of $A\snv$ are constant.

\begin{corollary}
Suppose $A$ is a unitary type-II matrix.  If $A$ is a spin model, the column sets of the matrices $I$, $A$ and $D_jA$ form a set of three mutually unbiased bases.
\end{corollary}

\begin{proof}
By the previous lemma, $A\inv D_jA =D_jAD_j\inv$.  The diagonal entries of $D_j$ have
norm 1, and so $D_j$ is unitary.  Hence $D_jAD_j\inv$ is a flat unitary matrix and therefore
$A\inv D_jA$ is flat and unitary.
\end{proof}

All of the known maximal sets of mutually unbiased bases are equivalent to a set of the form
\[
\{I, A, D_1A, \ldots, D_{n-1}A\},
\]
where each $D_i$ is diagonal and $A$ is the character table of the additive group of $GF(n)$ (which is type-II).

We consider one example of spin models.  Suppose $\th$ is a root of unity and let $W$ by the $n\times n$
matrix with rows and columns indexed by $0,1,\ldots,n-1$ and with $ij$-entry $\th^{(i-j)^2}$.  Then
\begin{align*}
(W^*W)_{r,s} &=\sum_{i=0}^{n-1} \th^{-(r-i)^2+(s-i)^2}\\
	&=\sum_{i=0}^{n-1} \th^{(s-r)(s+r-2i)}\\
	&=\th^{s^2-r^2}\sum_{i=0}^{n-1} \th^{2(r-s)i}.
\end{align*}
It follows that $W$ is type II if and only if $\th^2$ is a primitive $n$-th root of unity.
Clearly $W$ is flat.

Now
\[
(W_{r/s})_i  =\th^{(r-i)^2-(s-i)^2} =\th^{(r-s)(r+s-2i)} =\th^{r^2-s^2}\th^{2(r-s)i}
\]
and since $W$ is a circulant, it follows that $W$ is a spin model when $\th^2$ is a primitive
$n$-th root of unity.

All known examples of unitary spin models arise from character tables of finite abelian groups.
There are examples of non-unitary spin models---one due to Jaeger \cite{jae1} coming from the
Higman-Sims graphs and second family due to Nomura \cite{nom1} coming from Hadamard matrices.

\bibliographystyle{plain}
\bibliography{mub}


\end{document}